# Ultra-Broadband Dispersion-Manipulated Dielectric Metalenses by Nonlinear Dispersive Phase Compensation

Yueqiang Hu[1,2], Yuting Jiang[1], Yi Zhang[1], Jiajie Lai[1], Peng He[1], Xiangnian Ou[1], Ling Li[1] and Huigao Duan[1,2*]

[1]National Research Center for High-Efficiency Grinding, College of Mechanical and Vehicle Engineering, Hunan University, Changsha 410082, P.R. China

[2]Advanced Manufacturing Laboratory of Micro-Nano Optical Devices, Shenzhen Research Institute, Hunan University, Shenzhen, 518000, China

[*]Corresponding authors. Email: duanhg@hnu.edu.cn

**Abstract:**

Dispersion decomposes compound light into monochromatic components at different spatial locations, which needs to be eliminated in imaging but utilized in spectral detection. Metasurfaces provide a unique path to modulate the dispersion only by adjusting the structural parameters without changing the material as required for refractive elements. However, the common linear phase compensation does not conform to the dispersion characteristics of the meta-unit limiting dispersion modulation in broader wavelength bands, which is desired for ultra-broadband or multiband imaging. Here, we propose a nonlinear dispersive phase compensation method to design polarization-insensitive achromatic metalenses from 400 nm to 1000 nm constructed with single-layer high aspect ratio nanostructures. This band matches the response spectrum of a typical CMOS sensor for both visible and near-infrared imaging applications without additional lens replacement. Moreover, the capability of the method in achieving arbitrary dispersion modulation is demonstrated for applications such as chromatography imaging and spectral detection.

## Introduction

Dispersion is a fundamental property of materials, i.e., the refractive index of material with normal dispersion (e.g., glass) decreases with increasing wavelength. As a result, the prism deflects light at a longer wavelength by a smaller angle, and the focal lengths at a longer wavelength of refractive lenses are larger than for a shorter wavelength (i.e., chromatic aberration). The opposite is true for dispersion in diffractive optical elements. The manipulation of dispersion has been of wide interest and has numerous important applications. On the one hand, the chromatic aberration in lens imaging [1] and near-eye displays[2] severely degrades the image quality, thus chromatic aberration should be eliminated. On the other hand, increasing dispersion (hyper-dispersion) can enable higher resolution of spectrometer devices[3] and transmission capacity of wavelength division multiplexing in optical communications[4]. However, the current manipulation of dispersion requires the stacking of multiple components leading to bulky and complex systems, which are not conducive to the applications of miniaturized optical systems in consumer electronics, wearable devices, and miniature spectrometers.

Metasurfaces provide an attractive platform to design ultra-thin planar optical elements by constructing subwavelength scatters (metaunits) in a two-dimensional plane[5-7]. Combined with the multiparametric control capability of metaunits, a variety of elements with plentiful functionality, such as beam deflectors, metalenses[8-11], metaholograms[12-15] and complex beam generators[16-18], have been demonstrated. However, most of these components are heavily chromatic. Previous efforts have been done to achieve chromatic aberration elimination at discrete wavelengths through spatial multiplexing, such as interleaved metaunits[19-20] and stacking layers[21-24]. Meanwhile, with structural dispersion design freedom of different sub-wavelength structures, the dispersion regulation at discrete wavelengths[25], narrowband[26-27], and broadband[28-33] in different wavelength bands with a single-layer metalens can be realized through dispersive phase compensation. However, the current achromatic bandwidth is difficult to break further, especially for the short wavelengths such as near-infrared, visible or even ultraviolet band. First, the previous achromatic schemes were generally based on linear dispersion compensation, which is approximately valid in the narrow bandwidth. While when the wavelength is further reduced or the bandwidth is further increased, the structural dispersion relationship shows a strong nonlinearity leading to a notable increase in the compensation error. Second, the limited height of the subwavelength structures makes it difficult to cover a large dispersion compensation range in short wavelengths, hindering the realization of larger bandwidths or metalenses sizes.

In this work, we propose and demonstrate ultra-broadband dispersion-manipulated dielectric metalenses from 400 nm to 1000 nm by nonlinear dispersive phase compensation method. By constructing high-aspect-ratio $TiO_2$ nanostructures (height of 1000 nm and minimum width of 50 nm) with different cross-sectional shapes to form a large meta-atoms library, whose phase dispersion exhibits stronger nonlinear properties over a broader band. Then, the wavefront at each wavelength to precisely match with the arbitrary nonlinear dispersion of meta-atoms is independently constructed to achieve an ultra-broadband dispersion manipulation. The method is also utilized in the demonstration of the realization of customized dispersion (positive dispersion, super-negative dispersion, arbitrary dispersion). The operating band of the proposed metalenses is well matched to the response band of commonly used CMOS sensors for dual-band imaging in the visible and near-infrared. Moreover, the proposed method enables more accurate phase matching in dispersion modulation designs, making it possible to extend to ultra-wideband and multi-spectrum dispersion modulation applications.

**Results**

Achieving arbitrary manipulation of the dispersion of an optical element requires the design of independent phase profiles at all design wavelengths. The phase profiles in spectral space (phase dispersion) of the subwavelength structures, i.e. meta-units, are highly correlated with their cross-sectional shapes, providing new degrees of freedom to the dispersion manipulation with metasurfaces. Designing meta-units with the unique phase dispersion at specific locations of the element to meet the requirement of each wavelength is expected to achieve dispersion manipulation. In order to understand the structural dispersion mechanism of the dielectric subwavelength meta-units, we consider the meta-units as vertical waveguides, whose phase response can be expressed as

$$\varphi_{meta} = \frac{2\pi}{\lambda} n_{eff}(\lambda) H \quad (1)$$

where $H$ is the height of the meta-unit. $\lambda$ is the wavelength. $n_{eff}(\lambda)$ is the effective refractive index (ERI) related with the intrinsic refractive index and cross-section shape of the meta-unit. From Eq. (1), for a fixed height meta-unit, the phase is positively correlated with the wavenumber ($\frac{2\pi}{\lambda}$) and the exact relationship is determined by the ERI and is linear if ERI is independent of wavelength. However, according to the one-dimensional equivalent medium theory, the ERI generally varies with wavelength (see section 1 in the Supporting Information). The two-dimensional structures of meta-units are similar, as different waveguide modes lead to different ERI at different wavelengths. Eventually the phase dispersion will show nonlinear characteristics, which will be elaborated in detail subsequently. It can also be seen that the dispersion compensation range by the meta-units is proportional to the height of the structure.

Therefore higher structures help achieve wider bandwidth, larger area dispersion modulation designs, which require structures with larger aspect ratios.

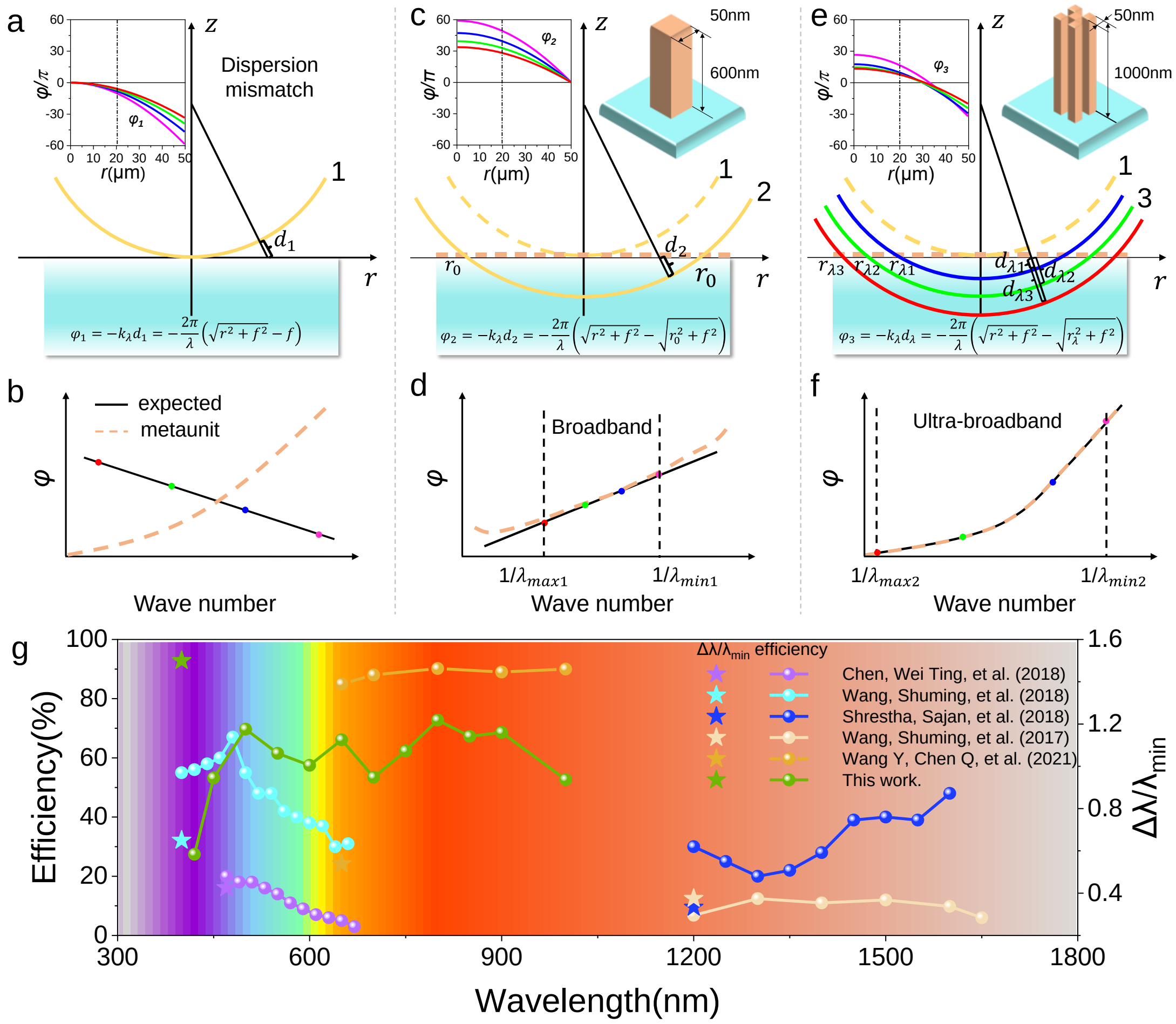

***Figure 1. The schematic of ultra-broadband dispersion-manipulated dielectric metalenses design.*** *(a, c, e) The schematic wavefronts of conventional lens, linear phase compensation scheme and nonlinear phase compensation scheme. The left embedded figure is the phase profiles of different wavelengths which are getting smaller from red to purple, the right embedded figure is a schematic diagram of nanostructures with different height that compose the metalens. (b, d, f) The comparison between the intrinsic phase dispersion of the meta-unit and constructed phase dispersion of traditional lens, linear phase compensation scheme and nonlinear phase compensation scheme respectively. (g) The efficiencies and relative bandwidth ($\Delta\lambda/\lambda_{min}$) of broadband achromatic metalenses in different works. This work achieves the*

*achromatic metalens with the largest absolute and relative bandwidth.*

A broadband achromatic metalens is the most typical form of dispersion manipulation. In order to focus all wavelengths of light to the same focal point, the wavefront constructive interferenced by each light path from the interface should be a hemisphere with the same spherical center. So the wavelength-dependent phase profile at the interface needs to be designed to construct such wavefront by accumulated with the propagating phase in free space, i.e., $\varphi = -k_\lambda d$, where $d$ is the physical distance between the wavefront and the interface at a specific location. As shown in Figure 1a, for the conventional lens wavefront, $d_1$ is intuitively designed to construct a phase distribution of $\varphi_1 = -\frac{2\pi}{\lambda}\left(\sqrt{r^2+f^2}-f\right)$, where $f$ is the focal length. The dispersion relationship is shown in the embedded figure where wavelength of red line is the largest. For a certain position of the lens, the phase is negatively correlated with the wavenumber as shown in Figure 1b. Therefore, it is clear from Eq. (1) that phase compensation cannot be achieved with the meta-units. The scheme commonly used in the previous studies is shown in Figure 1c[29, 31, 34], where the wavefront is expanded outward to the reference position $r_0$ and the phase distribution is obtained as $\varphi_2 = -\frac{2\pi}{\lambda}\left(\sqrt{r^2+f^2}-\sqrt{r_0^2+f^2}\right)$. The dispersion relationship in the embedded figure shows that the phase profiles are translated upward and intersected at position $r_0$. In this way, the phase dispersion sign is flipped at positions smaller than $r_0$ as shown in Figure 1d, providing the possibility of phase compensation by meta-units. However, since $r_0$ is a constant resulting linear relationship between the constructed phase and the wavenumber, the chromatic aberration can be eliminated only in a narrow bandwidth with an approximate linear fit with the meta-units' dispersion (Figure 1d). It will introduce large error for wider bandwidth and the meta-units library selection is limited.

To better match with the nonlinear structural dispersion, we propose a new phase dispersion construction method as shown in Figure 1e which can expressed as

$$\varphi_3 = -kd_\lambda = -\frac{2\pi}{\lambda}\left(\sqrt{r^2 + f^2} - \sqrt{r_\lambda^2 + f^2}\right) \quad (2)$$

$r_\lambda$ introduced here is a wavelength dependent value. This means that different $d_\lambda$ are achieved for different wavefront translations at each wavelength, resulting in the same focus at each wavelength even though the wavefront do not overlap. So the dispersion relationship in Fig. 1f can be constructed to exactly match the structural dispersion, thus achieving chromatic aberration eliminated in the ultra-broadband range. Further, the wavelength-dependent phase required for each position can be decomposed from Eq. (2) into two parts $\varphi(r,\lambda) = \varphi(r,\lambda_{\max}) + \Delta\varphi(r,\lambda)$, where the first term is the phase at the maximum wavelength of the achromatic band $\lambda_{\max}$ (reference phase) and the second term is the phase difference between wavelength $\lambda$ and $\lambda_{\max}$. Therefore, in the design, we need to construct the appropriate $r_\lambda$ so that the appropriate meta-unit chosen for each position can satisfy the phase and phase difference. Figure 1g shows the performance of the ultra-broadband achromatic metalens achieveed in this work compared to previous ones. This work have achieved the largest absolute and relative achromatic bandwidths ($\frac{\lambda}{\lambda_{min}}$, for uniform evaluation of bandwidths at different bands). And relatively high focusing efficiency is maintained in this band. The design and experimental results are described in detail below.

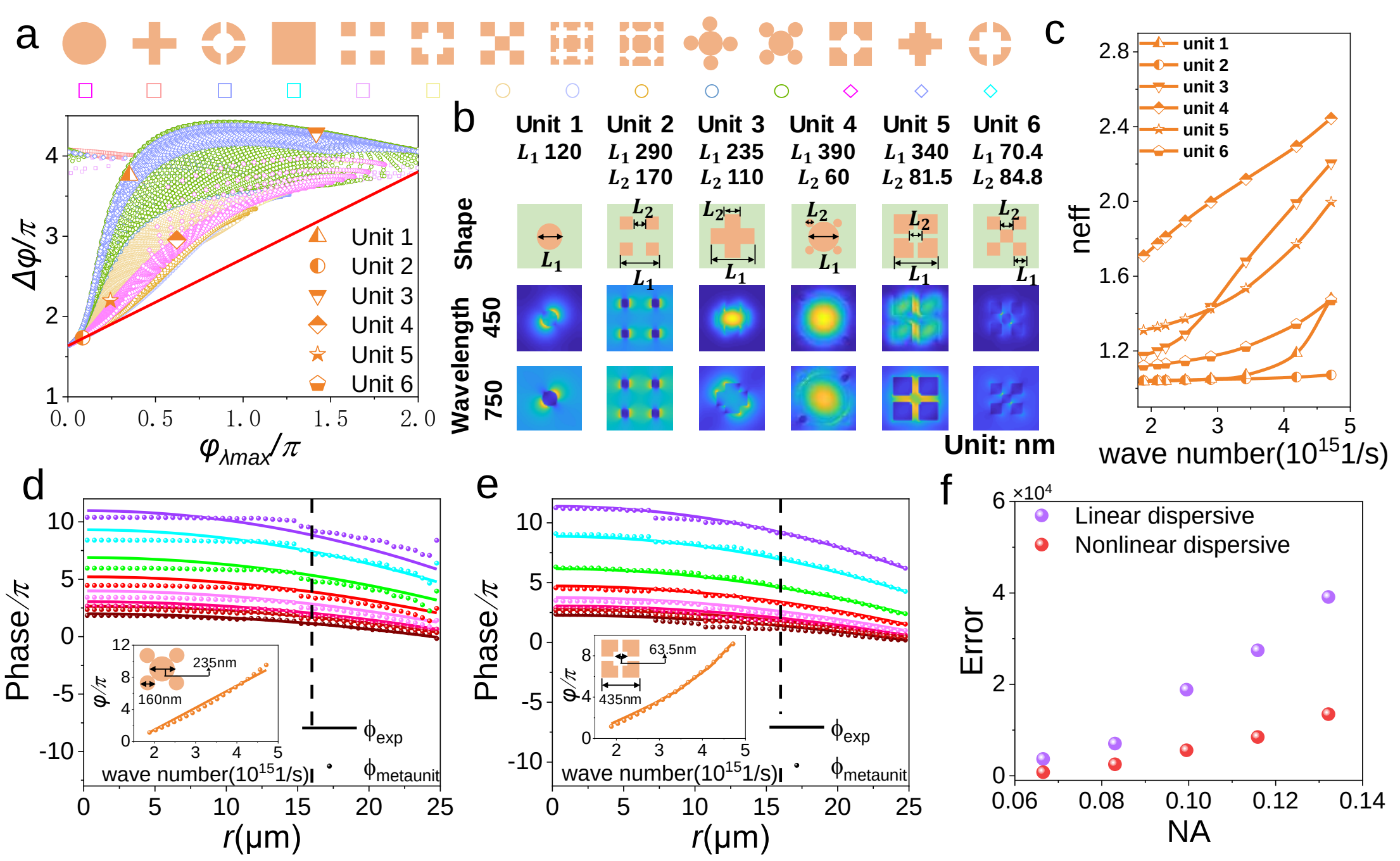

***Figure 2. The design process of ultra-broadband dispersion-manipulated dielectric metalenses.*** *(a) The schematic diagram of the cross-section of 14 types of nanostructures and the "phase-phase dispersion" library calculated at 550 nm wavelength. The red line is the effective medium line representing the theoretical limit with minimal dispersion (no structural dispersion). (b) The shapes and parameters of six meta-units and the waveguide modes at two different wavelengths. (c) The relationship between the ERI of six meta-units and the wave number. (d, e) The matched results with linear and nonlinear dispersive compensation methods in the radius dimension for different wavelengths of the metalens with a radius of 25μm and NA of 0.083, where the solid lines are the constructed wavefront phase profiles and the scatters are the phases of the matched meta-units. The embedding image is the phase dispersion figure of the structure at the position of 15.75μm radius. (f) The total phase errors at all wavelengths of the metalenses with different NA. The purple point and the red point are the errors of linear and nonlinear dispersive compensation methods.*

To acquire a sufficient number of meta-units to achieve the desired dispersion compensation, a large database of polarization-insensitive dielectric meta-units with different cross-sectional shapes and parameters should be built. $TiO_2$ material is chosen to achieve the visible to near-IR design due to its optical constant measurement result in Figure S2, which shows a large refractive index with negligible absorption in this band. The meta-units are modeled as dielectric waveguides in transmission mode with wavelength-dependent effective refractive indices $n_{eff}$ which calculated by eigenmode analysis (see Methods). The phase is then obtained by Eq. (1), where the height of the structure is set as 1000 nm to achieve a larger range of dispersion. We designed 14 types of cross-sectional shapes considering fabrication constraints, and parametric scans were performed under each shape. Since the key parameters of the meta-units are the reference phase and phase difference, all databases are plotted in the "phase-phase dispersion" space, as shown in Fig. 2a for the wavelength of 550 nm (maximum wavelength is 1000 nm). The dispersion results for other wavelengths and 600 nm-height structures in the section 3 in the Supporting Information show that the greater the range of dispersion can be compensated with higher structures. Each of these points represents the meta-unit of a particular parameter under a shape. It can be seen that the subclasses of each shape fill a different region, representing that they have different structural dispersions. By comparing with the effective medium line representing the theoretical limit with minimal dispersion (no structural dispersion), our database fully covers the possible structural dispersion regions densely, thus ensuring the completeness of our achromatic design. Six meta-units with different phase dispersions due to different shapes and different sizes are selected in Fig. 2a to study their structural dispersion properties. Figure 2b shows the shapes and parameters of these six meta-units and the waveguide modes at two different wavelengths. It can be seen that the waveguide modes of the meta-units differ significantly at different wavelengths, which results in different ERIs at different wavenumbers in Figure 2c. Thus, from Eq. (1), it is proved that the meta-units have nonlinear phase dispersions due to ERIs varying with wavelengths (see Figure S5). Meanwhile, significantly different ERIs variations can be found for different meta-units, also providing great freedom to compensate for the phase difference at different locations.

Using the established database, for comparison, we designed ultra-broadband achromatic metalenses with different numerical apertures (NA) in the 400 nm to 1000 nm band by linear and nonlinear dispersive phase compensation methods, respectively. The chosen band covers the spectral response range of a general CMOS image sensor and enables multi-band imaging in the visible and near infrared. Figure 2d, e show the matching results in the radius dimension for different wavelengths of the metalens with a radius of 25 μm and NA of 0.083, where the solid lines are the constructed wavefront phase profiles and the scatters are the phases of the matched meta-units. It can be seen that the nonlinear dispersive phase compensation method is better matched than the linear one at most wavelengths. And the phase dispersion plots at selected specific locations (embedding plots) shows the reason, i.e., the intrinsic phase dispersion of the selected meta-unit matches better with the constructed nonlinear dispersive phase dispersion in the large operation band. The total phase errors at all wavelengths of the metalenses with different NA in Figure 2f further illustrate the advantage of nonlinear dispersive phase compensation methods over linear one (see section 5 in the Supporting Information for detailed matching results). As the NA increases, the error of the linear method grows significantly faster than that of the nonlinear method showing that the advantage of the nonlinear method is more obvious under lager NA.

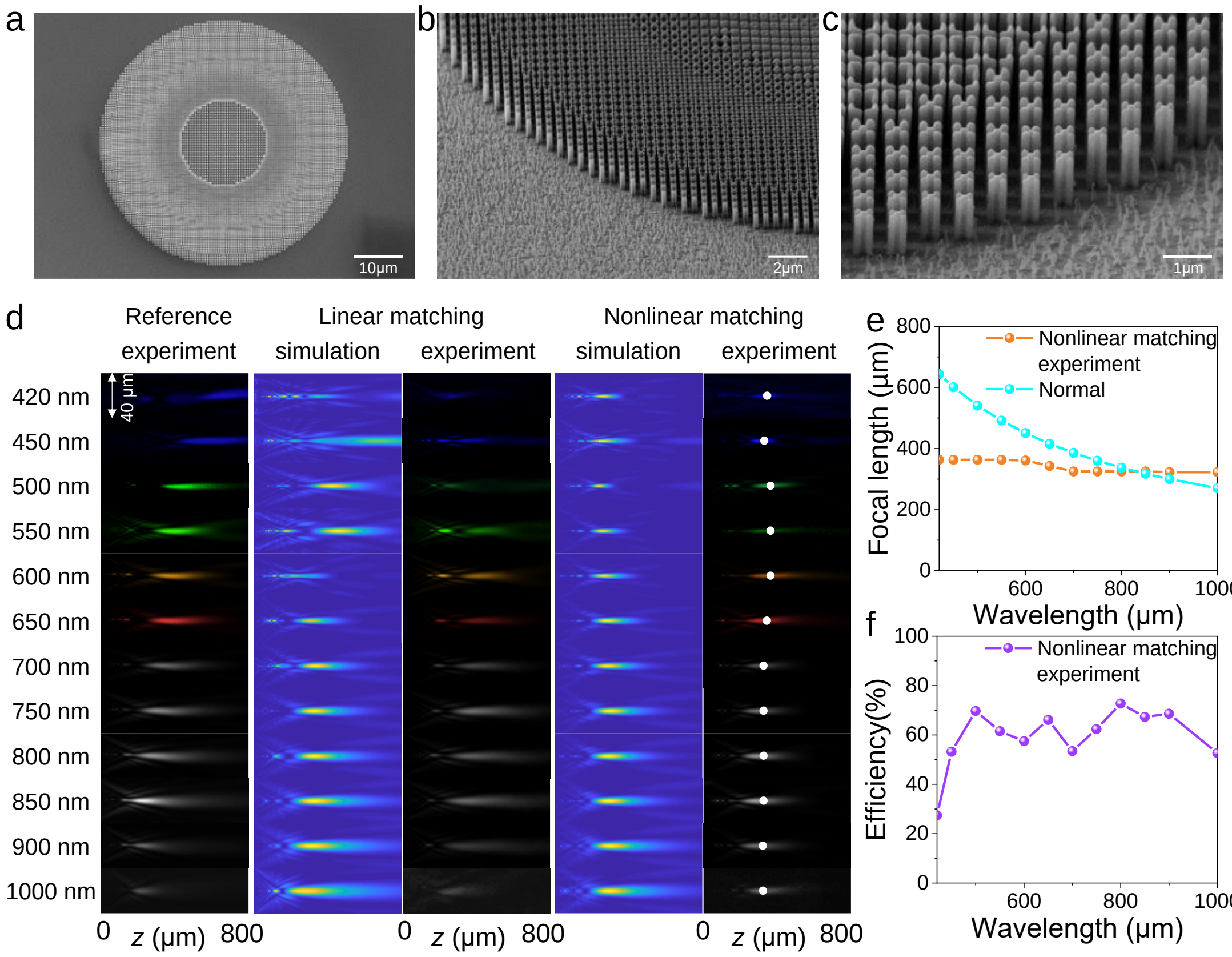

***Figure 3． Experimental results of ultra-broadband achromatic metalens with NA = 0.083.*** *(a-c) Three magnification scanning electron microscope (SEM) images of the fabricated achromatic metalens with NA = 0.083. a. Scale bar: 10μm. b. Scale bar: 2μm. c. Scale bar: 1μm. (d) The vector simulation and experimental intensity distribution along the propagation direction (z axis) of the three metalenses. From the left to the right are experimental results of the reference group, the simulation results and experimental results of the linear matched group, and the simulation results and experimental results of the nonlinear matched group. (e) Focal length distribution at different wavelengths. The orange line is the focal length of the nonlinear matched metalens, and the blue line is the normal negative dispersion reference curve. (f) The measured focusing efficiency of the nonlinear matched metalens.*

To achieve high quality processing of metalenses consisting of high aspect ratio meta-units, we developed a conformal filling method based on electron beam lithography (EBL), atomic layer deposition (ALD), as described detailed in the method section and Figure S8. Figures 3a, b, c show the scanning electron microscopy (SEM) images of the fabricated achromatic metalens with NA = 0.083 at different magnifications, respectively. It can be seen that the processes ensure the steepness of the nanostructures well and enable the aspect ratio larger than 20. A single wavelength designed metalens for reference and a metalens designed by linear dispersive phase compensation method were also fabricated. The fabricated metalenses were measured using the optical experimental setup in Section S7, Supporting Information to characterize the focusing and imaging performance. Figure 4d shows the vector simulation and experimental intensity distribution along the propagation direction ($z$ axis) of the three metalenses. First, the experimental results of the reference group exhibit the normal negative dispersion characteristics of the single-wavelength designed metalens, i.e., the focal length decreases as the wavelength becomes larger. The linear matching metalens can achieve achromatic focusing in a certain bandwidth (e.g., 700-900 nm), but the large matching error makes its focusing poor in the visible range. While the metalens designed with nonlinear dispersive phase compensation can achieve good chromatic aberration elimination from 400 to 1000 nm. The shift of the focus in the experiment is due to the processing error.  Figure 4e, f show the focal length and measured focusing efficiency of the nonlinear matching metalens, where the focusing efficiency is calculated by the intensity within three times the full width half height (FWHM) at the focus plane divided by the whole intensity at the focus plane. It can be seen that the designed metalens achieves achromatic focusing while also having high focusing efficiency in the operation band. This is due to the small matching error and the high transmission of $TiO_2$ throughout the visible to near-infrared..

Figure 4a, c show the focal spot profiles of the nonlinear matching achromatic

metalens in the visible and near-infrared range, respectively. The imaging performance of a standard United States Air Force resolution target from the metalens are shown in Figure 4b, d under various incoherent illumination lights with a bandwidth of 20 nm. In the measurements, the target and the image plane were respectively set as a fixed plane for all the wavelengths to evaluate the achromatic performance of the metalens. The imaging of element 2 in group 5 on the resolution target are measured. It can be seen that high contrast imaging can be achieved at most wavelengths demonstrating the function of the metalens in ultra-broadband achromatic imaging. A decrease in focusing efficiency results in a decrease in contrast at the blue wavelengths. Note that the different resolution of the visible and near-infrared band detectors lead to differences in the imaging under the two bands.

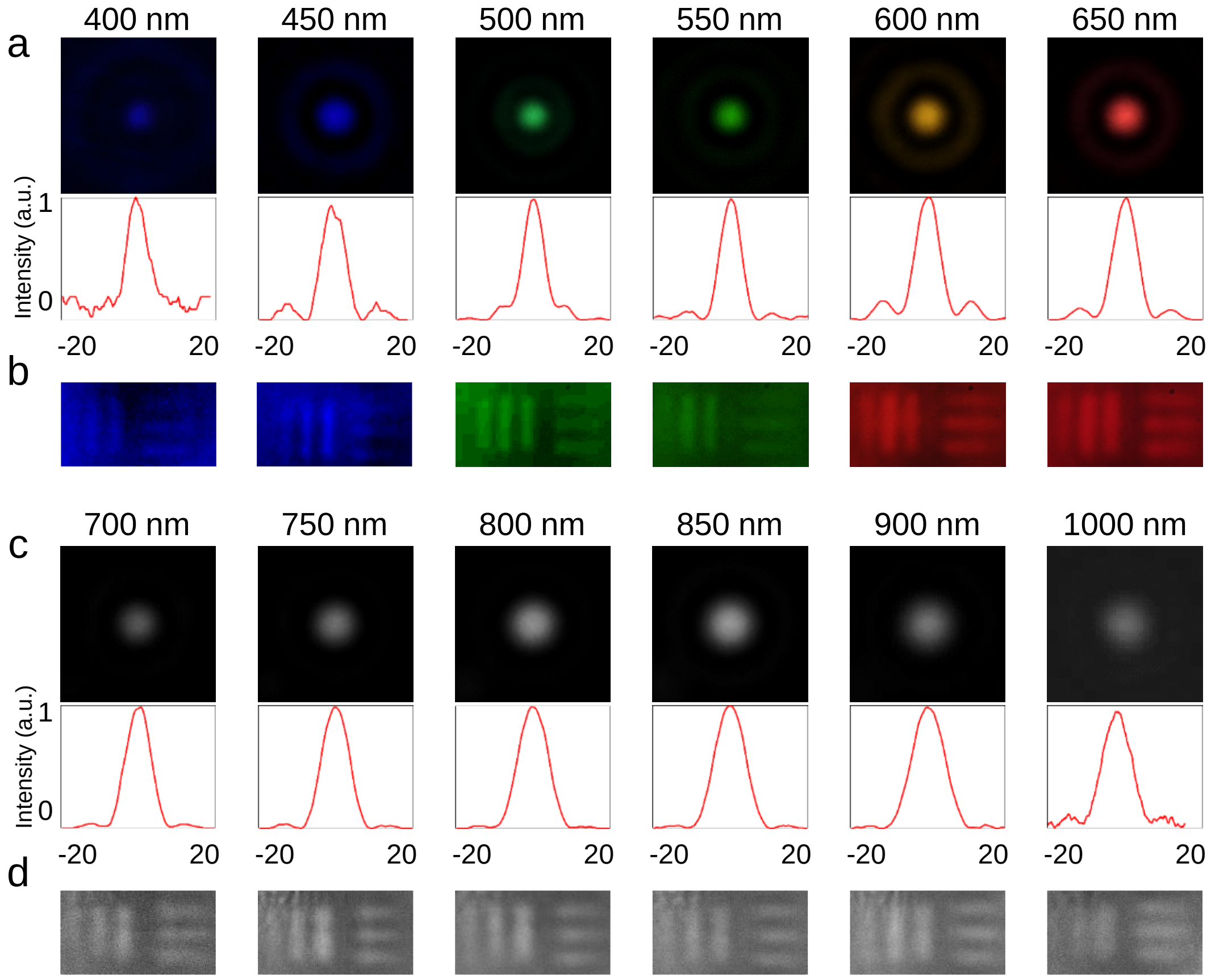

***Figure 4．The focus characterization and imaging performance of ultra-wideband achromatic metalens.*** *(a, c) Focal spot profiles and normalized intensity profiles for*

*various wavelengths. (b, d) Images of element 2 in group 5 on the 1951 United States Air Force resolution target formed by the achromatic metalens.*

To further demonstrate the applicability of nonlinear matching methods for ultra-broadband dispersion manipulation, we designed and fabricated three customized metalenses to achieve enhanced negative dispersion, positive dispersion, and arbitrary dispersion manipulation. Customized dispersion design is achieved by constructing the wavefront phase at each wavelength for the desired focus, where the nonlinear matching scheme provides great freedom to design each individual wavefront. The three metalenses have a diameter of 50 μm and a focal length of 320 μm at 650 nm. Figure 5a, b, c show the measured intensity distribution along the propagation direction for these three metalenses in 400~1000 nm band. All the three metalenses achieve good focus at each wavelength but with significantly different chromatic aberrations. Figure d, e, f show the focal length statistics at all wavelengths, where the blue line is the normal negative dispersion reference curve. It can be seen that the three designed metalenses achieve enhanced negative dispersion, positive dispersion and arbitrary dispersion modulation.

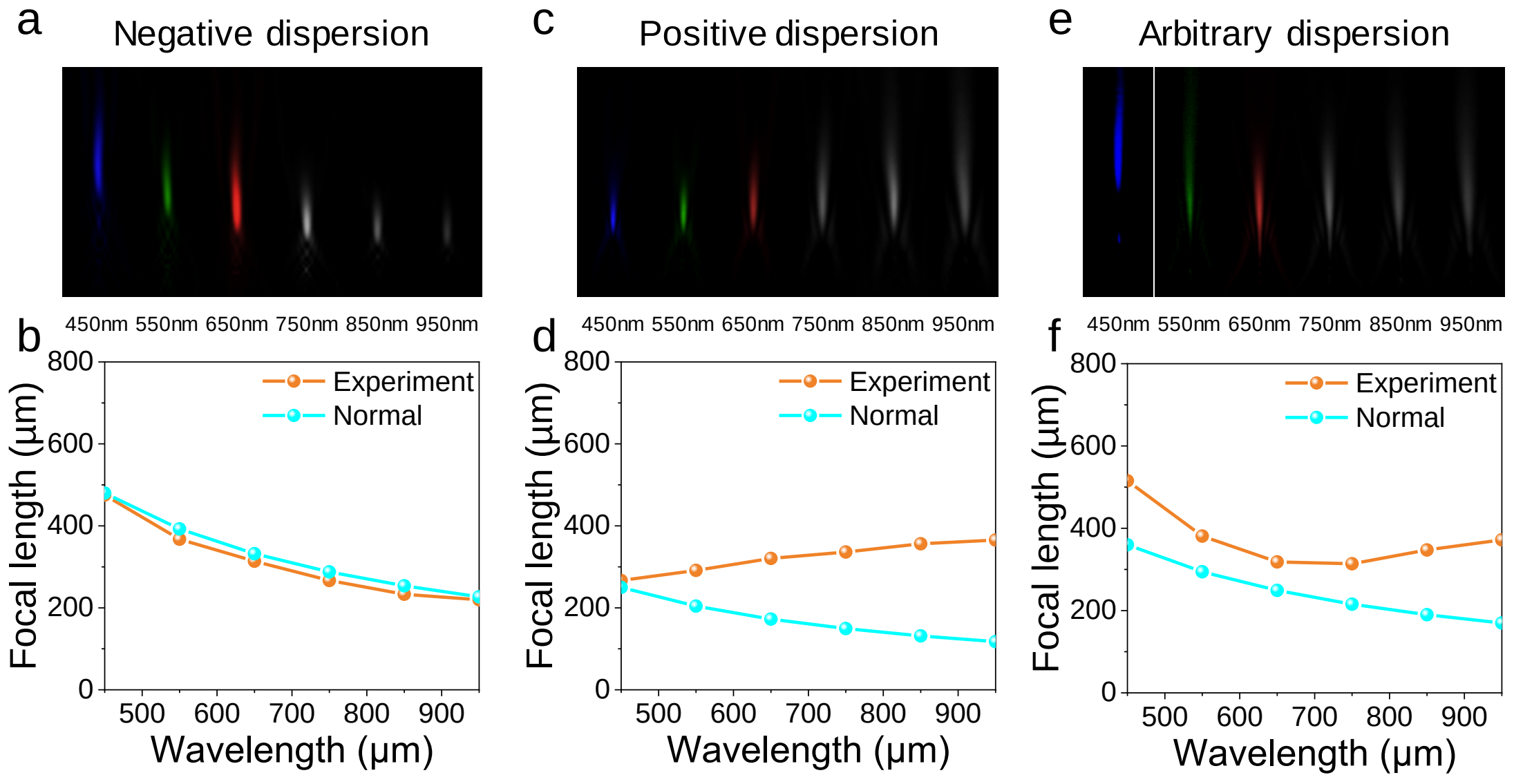

***Figure 5．Experimental results of arbitrary dispersion control metalens.*** *(a-c)*

*Experimental light intensity profiles for the enhanced negative dispersion metalens, positive dispersion metalens and arbitrary dispersion metalens respectively with NA = 0.083 at various incident wavelengths. (d-f) The orange line is the focal length statistics at all wavelengths, and the blue line is the normal negative dispersion reference curve.*

**Discussion**

Our proposed nonlinear dispersive phase compensation method provides a powerful tool for dispersion modulation in ultra-broadband enabling chromatic aberration cancellation in even larger bandwidths, such as from near ultraviolet to near-infrared (see Figure S12). Moreover, this nonlinear matching scheme is convenient for applying optimization algorithms such as particle swarm algorithms, genetic algorithms to optimize $r_\lambda$ at each wavelength instead of using the lookup table method. The customized dispersion manipulation capability can be applied in various applications such as color holography[35], spectral detection[3], wave division multiplexing optical communication. This method still does not break the limit of the maximum dispersion range of the nanostructures, i.e., realizing achromatic metalenses with larger diameters and higher NA requires a larger range of phase dispersion supported by increasing the refractive index of the material and the height of the meta-units. But this scheme provides a possible idea to break through this limit at discrete wavelengths by wrapping all the phase profiles to $2\pi$ range and then optimizing $r_\lambda$ for each wavelength. Since unavoidable processing errors can lead to changes in lens performance, we analyze the effects of processing errors in the supplementary material.

**Conclusion**

In summary, we proposed a nonlinear dispersive phase compensation scheme that better matches the intrinsic phase dispersion response of nanostructures to address the obstacle of large errors in dispersion manipulation designs with linear matching at wider bandwidths. With this scheme, we have demonstrated ultra-broadband achromatic and customized dispersive dielectric metalenses spanning the visible to near-infrared band from 400 to 1000 nm composed of carefully fabricated high aspect ratio nanostructures. This is the widest bandwidth of dispersion manipulation achieved by a metalens so far, which covers the response band of general CMOS image sensors enabling the achromatic imaging in both day and night environments with the same extremely miniaturized optical system. In addition, this scheme provides great freedom to achieve various types of customized dispersion modulation for applications such as color holography, and spectral detection.

**Methods**

**Numerical simulation.** The ERIs of 14 kinds of meta-units are simulated by the Lumerical MODE Solutions. The period of meta-atoms were set as 500 nm. Considering the constraints of experimental conditions and period size, we set the minimum and maximum size constraints of the nanofins to be 50 nm and 450 nm, respectively. Nanostructures with different cross-sectional shapes are shown in Figure 2a. For the simulation, the boundary conditions of the simulation are set to periodic boundary conditions. The refractive index of the $TiO_2$ was the measurement result by ellipsometer. We used eigenmodes to analyze and calculate nanostructures with different wavelengths and different cross-sectional shapes, and obtain the equivalent refractive index $n_{eff}$.

**Device fabrication:** First, a 1000-nm-thick polymethyl methacrylate (PMMA) electron-beam resist layer was spin coated at 2000 rpm on the transparent glass substrate with ITO film layer and bake on a hot plate for 4 min at 180 ℃. Then, the sample was exposed by electron-beam lithography (EBL) with a 100-KV voltage and a beam current of 200 pA. Subsequently, we put the exposed sample in a mixed solution of isopropanol and methyl isobutyl ketone (IPA: MIBK = 3:1) for 3 minutes, and then fixed it in the IPA solution for 1 minute at room temperature. Later, we used the atomic layer deposition (ALD) system to fill the exposed area with 230 nm $TiO_2$. After this process, there will be a layer of 230 nm $TiO_2$ on the top of the entire sample, we removed it by ion beam etching (IBE) in the next process. After removing the $TiO_2$ on the top layer, we used reactive ion etching (RIE) to remove the resist. Finally, the $TiO_2$ nanostructures with a high aspect ratio are obtained.

**Optical characterization.** In order to verify the performance of ultra-broadband achromatic metalens and arbitrary dispersion control metalenses, we design two optical setups to characterize the focal length and imaging effect, respectively. Details of the optical experimental setup for characterizing the ultra-broadband achromatic metalens are shown in supplementary information.

**Acknowledgments**

We acknowledge the financial support by the National Natural Science Foundation of China (Grant No. 52005175, 5211101255), Natural Science Foundation of Hunan Province of China (Grant No. 2020JJ5059) and Shenzhen Science and Technology Program (Grant No. RCBS20200714114855118).

**Author contributions:**

Y.H. proposed the idea. Y.H., Y.J. and Y.Z. conceived and carried out the design and simulation. Y.H., Y.J. and J.L. fabricated the samples. Y.H., Y.J., P.H. and X.O. conceived and performed the measurements. Y.H., H.D., Y.J. and L.L discussed the results and co-wrote the manuscript. H.D. supervised the overall project. All the authors discussed the results and commented on the manuscript.

**Competing interests:** The authors declare that they have no competing interests.